\documentclass[prl,final,showpacs,floatfix,superscriptaddress,nofootinbib]{revtex4}

\usepackage[utf8]{inputenc}
\usepackage{amsfonts}
\usepackage{amsmath}
\usepackage{amssymb}
\usepackage{graphicx}
\usepackage{nicefrac}
\usepackage{slashed}
\usepackage{bigstrut}
\usepackage{morefloats}
\usepackage{multirow}
\usepackage{longtable}
\usepackage{dcolumn}
\usepackage{placeins}
\usepackage{soul}
\usepackage{hyperref}

\hypersetup{
        colorlinks      =true,
        urlcolor        =blue,
        linkcolor       =blue,
        citecolor       =blue,
        }
        
\allowdisplaybreaks[1]

\providecommand{\U}[1]{\protect\rule{.1in}{.1in}}

\begin{document}

\title{Nuclear binding near a quantum phase transition}

\author{Serdar~Elhatisari}
\affiliation{Helmholtz-Institut f\"ur Strahlen- und
             Kernphysik and Bethe Center for Theoretical Physics,
             Universit\"at Bonn,  D-53115 Bonn, Germany}
             
\author{Ning~Li} 
\affiliation{Institute~for~Advanced~Simulation, Institut~f\"{u}r~Kernphysik,
and
J\"{u}lich~Center~for~Hadron~Physics,~Forschungszentrum~J\"{u}lich,
D-52425~J\"{u}lich, Germany}

\author{Alexander~Rokash}
\affiliation{Institut~f\"{u}r~Theoretische~Physik~II,~Ruhr-Universit\"{a}t~Bochum,
D-44870~Bochum,~Germany}

\author{Jose~Manuel~Alarc\'on} 
\affiliation{Helmholtz-Institut f\"ur Strahlen- und
             Kernphysik and Bethe Center for Theoretical Physics,
             Universit\"at Bonn,  D-53115 Bonn, Germany}  
               
\author{Dechuan~Du}   
\affiliation{Institute~for~Advanced~Simulation, Institut~f\"{u}r~Kernphysik,
and
J\"{u}lich~Center~for~Hadron~Physics,~Forschungszentrum~J\"{u}lich,
D-52425~J\"{u}lich, Germany} 
       
\author{Nico~Klein}
\affiliation{Helmholtz-Institut f\"ur Strahlen- und
             Kernphysik and Bethe Center for Theoretical Physics,
             Universit\"at Bonn,  D-53115 Bonn, Germany} 

\author{Bing-nan~Lu}
\affiliation{Institute~for~Advanced~Simulation, Institut~f\"{u}r~Kernphysik,
and
J\"{u}lich~Center~for~Hadron~Physics,~Forschungszentrum~J\"{u}lich,
D-52425~J\"{u}lich, Germany}

\author{Ulf-G.~Mei{\ss}ner}  
\affiliation{Helmholtz-Institut f\"ur Strahlen- und
             Kernphysik and Bethe Center for Theoretical Physics,
             Universit\"at Bonn,  D-53115 Bonn, Germany}  
\affiliation{Institute~for~Advanced~Simulation, Institut~f\"{u}r~Kernphysik,
and
J\"{u}lich~Center~for~Hadron~Physics,~Forschungszentrum~J\"{u}lich,
D-52425~J\"{u}lich, Germany}     
\affiliation{JARA~-~High~Performance~Computing, Forschungszentrum~J\"{u}lich,
D-52425 J\"{u}lich,~Germany}        

\author{Evgeny~Epelbaum}
\affiliation{Institut~f\"{u}r~Theoretische~Physik~II,~Ruhr-Universit\"{a}t~Bochum,
D-44870~Bochum,~Germany}

\author{Hermann~Krebs}
\affiliation{Institut~f\"{u}r~Theoretische~Physik~II,~Ruhr-Universit\"{a}t~Bochum,
D-44870~Bochum,~Germany}

\author{Timo~A.~L\"{a}hde}
\affiliation{Institute~for~Advanced~Simulation, Institut~f\"{u}r~Kernphysik,
and
J\"{u}lich~Center~for~Hadron~Physics,~Forschungszentrum~J\"{u}lich,
D-52425~J\"{u}lich, Germany}

\author{Dean~Lee}
\affiliation{Department~of~Physics, North~Carolina~State~University, Raleigh,
NC~27695, USA}

\author{Gautam~Rupak}
\affiliation{Department~of~Physics~and~Astronomy and HPC$^2$ Center for Computational
Sciences, Mississippi~State~University,
Mississippi State, MS~39762, USA}

\begin{abstract}
  How do protons and neutrons bind to form nuclei?  This is the central question
of \textit{ab initio} nuclear structure theory. While the answer may seem
as simple as the fact that nuclear forces are attractive, the full story
is more complex and interesting. In this work we present numerical evidence
from \textit{ab initio} lattice simulations showing that nature is near a
quantum phase transition, a zero-temperature transition driven by quantum
fluctuations.  Using lattice effective field theory, we perform Monte Carlo
simulations for systems with up to twenty nucleons.  For even and equal numbers
of protons and neutrons, we discover a first-order transition at zero temperature
from a Bose-condensed gas of alpha particles ($^4$He nuclei) to a nuclear
liquid. Whether one has an alpha-particle gas or nuclear liquid is determined
by the strength of the alpha-alpha interactions, and we show that the alpha-alpha
interactions depend on the strength and locality of the nucleon-nucleon interactions.
This insight should be useful in improving calculations
of nuclear structure and important astrophysical reactions involving alpha
capture on nuclei. Our findings also provide a tool to probe the structure
of alpha cluster states such
as the Hoyle state \cite{Funaki:2003af,Chernykh:2007a,Epelbaum:2011md,Zimmerman:2011,Epelbaum:2012qn,Dreyfuss:2012us}
responsible for the production of carbon in red giant stars and point to
a connection between nuclear states and the universal physics of bosons at
large scattering length  \cite{Efimov:1971a,Braaten:2004a}.
\end{abstract}

\flushbottom

\maketitle


There have been  significant recent advances in \textit{ab initio}  nuclear
structure theory using a variety of different \textit{}methods 
\cite{Roth:2011ar,Hergert:2012nb,Lahde:2013uqa,Soma:2013ona,Carlson:2014vla,Simonis:2015vja,Hagen:2015yea}.
Much of the progress has been driven by computational advances, but we also
have a better conceptual understanding of how nuclear forces impact nuclear
structure.  A key tool in making this connection is chiral effective
field theory, which organizes the low-energy nuclear interactions
of protons and neutrons according to powers of momenta and factors of the
pion mass.
   The most important interactions are included at leading order (LO), the
next largest contributions appear at next-to-leading order (NLO), and then
next-to-next-to-leading order (NNLO) and so on.
See Ref.~\cite{Epelbaum:2008ga}
for a
recent review of chiral effective field theory.
While the progress in \textit{ab initio}
nuclear theory has been impressive, there are gaps in our understanding of
the connection between nuclear forces and nuclear structure. In this letter
we discover an unexpected twist in the story of how nucleons self-assemble
into nuclei.  In order
to make our calculations transparent and reproducible by others, we  remove
all non-essential complications from our discussion. For this purpose, we
present  lattice Monte Carlo simulation results using lattice interactions
at LO in chiral effective field theory, together with Coulomb interactions
between protons. In the lattice calculations discussed here we use a spatial
lattice spacing of 1.97 fm and time lattice spacing of 1.32 fm.  We are using
natural units where the reduced Planck constant $\hbar$ and the speed of
light
$c$ equal 1. 

Our starting point is two lattice interactions A and B at leading order in
chiral effective field theory which are by design similar to each other and
tuned to experimental low-energy nucleon-nucleon scattering phase shifts.
The details of these interactions and scattering phase shifts are presented
in Supplemental Materials at [URL will be inserted by publisher], but we note some important points here. The interactions appear
at LO in chiral effective field theory and consist of short-range interactions
as well as the potential energy due to the exchange of a pion.  As the short-range
interactions are not truly point-like, they are actually what we call improved
LO\ interactions. We write the nucleon-nucleon interactions as $V_{\rm
A}({\bf r'},{\bf r})$ and $V_{\rm
B}({\bf r'},{\bf r})$, where ${\bf r}$ is the spatial
separation of the two incoming nucleons and ${\bf r'}$ is the spatial
separation of the two outgoing nucleons.
The short-range interactions in $V_{\rm
A}({\bf r'},{\bf r})$ consist of nonlocal terms, which means that ${\bf r'}$
and ${\bf r}$ are in general different.
In contrast, the short-range interactions in $V_{\rm
B}({\bf r'},{\bf r})$ include nonlocal terms and also local terms where ${\bf
r'}$ and ${\bf r}$ are fixed to be equal.  The main difference between interactions
A and B is the degree of locality of the short-range interactions.  Another
difference is that there are extra parameters contained in interaction B,
and these are used to reproduce $S$-wave scattering for two alpha particles.

We have used auxiliary-field Monte Carlo simulations to calculate nuclear
ground state energies. In Table~\ref{nuclei} we present the ground state
energies of ${^3}$H, ${^3}$He, ${^4}$He, ${^8}$Be, $^{12}$C,
$^{16}$O, $^{20}$Ne for interactions
A and B.  While we use the notation meant
for bound
nuclei, in some cases
the nuclear ground state is an unbound continuum state in our finite periodic
box.  We do not stabilize against decay to alpha particles.
 In fact, for the case of interaction A, all of the ground states
in Table 1 are multi-alpha states.  Details about the size of the box and the initial states used in the Monte Carlo simulations are provided in Supplemental Materials at [URL will be inserted by publisher].   The nuclei ${^4}$He, ${^8}$Be, $^{12}$C,
$^{16}$O, $^{20}$Ne are alpha-like nuclei with even and equal numbers of
protons and neutrons.  We show results at leading order (LO) and leading
order with Coulomb interactions between protons
(LO + Coulomb), as well as the comparison with experimental data.  All energies
are in units of MeV.  The lattice volume is taken large enough so that the
finite-volume energy correction is less than 1\% in relative error. The LO
+ Coulomb results for interaction B are in good agreement with experimental
results, better overall than the NNLO
results in Ref.~\cite{Lahde:2013uqa}.  However there
is significant underbinding for interaction A with increasing nucleon number.
 For interaction A, it is illuminating to compute the ratio of the LO energy
for each of the alpha-like nuclei to that of the alpha particle.  For $^8$Be
the ratio is 1.997(6), for $^{12}$C the ratio is 3.00(1), for $^{16}$O it
is 4.00(2), and for $^{20}$Ne we have 5.03(3).  These simple integer ratios
indicate that the ground state for interaction A in each case is a weakly-interacting
Bose gas of alpha particles.  This interpretation is also confirmed by calculations
of two-nucleon spatial correlations and local four-nucleon correlations.

\begin{table*} 
\caption{Ground state energies of ${^3}$H, ${^3}$He, ${^4}$He, ${^8}$Be,
$^{12}$C, $^{16}$O, $^{20}$Ne
for interactions A and B.  We show LO results, LO + Coulomb results, and
experimental data.  All energies are
in units of MeV.  The error bars denote one standard deviation errors.\bigskip}
\centering{}%
\begin{tabular}{c|c|c|c|c|c}
\hline 
Nucleus & A (LO) & B (LO) & A (LO + Coulomb) & B (LO + Coulomb) & Experiment
\tabularnewline
\hline 
$^{3}$H & $-7.82(5)$ & $-7.78(12)$ & $-7.82(5)$ & $-7.78(12)$ & $-8.482$
\tabularnewline
$^{3}$He & $-7.82(5)$ & $-7.78(12)$ & $-7.08(5)$ & $-7.09(12)$ & $-7.718$
\tabularnewline
$^{4}$He & $-29.36(4)$ & $-29.19(6)$ & $-28.62(4)$ & $-28.45(6)$ & $-28.296$
\tabularnewline
$^{8}$Be & $-58.61(14)$ & $-59.73(6)$ & $-56.51(14)$ & $-57.29(7)$ & $-56.591$
\tabularnewline
$^{12}$C & $-88.2(3)$ & $-95.0(5)$ & $-84.0(3)$ & $-89.9(5)$ & $-92.162$
\tabularnewline
$^{16}$O & $-117.5(6)$ & $-135.4(7)$ & $-110.5(6)$ & $-126.0(7)$ & $-127.619$
\tabularnewline
$^{20}$Ne & $-148(1)$ & $-178(1)$ & $-137(1)$ & $-164(1)$ & $-160.645$ 
\tabularnewline
\hline
\end{tabular}
\label{nuclei}
\end{table*}

\begin{figure}[!ht]
\centering
\caption {Alpha-alpha $S$-wave scattering. We plot $S$-wave phase shifts
$\delta_0$ for alpha-alpha scattering for interactions A and B versus laboratory
energy. We show LO results for interaction A (green triangles), LO + Coulomb
for A (orange diamonds), LO results for B (blue circles), and LO + Coulomb
results for B (red squares).  The phase shift analysis of experimental data
are shown with black asterisks \cite{Afzal:1969}.  The theoretical error
bars indicate one standard deviation uncertainty due to Monte Carlo errors
and the extrapolation to infinite number of time steps. \bigskip }
\includegraphics[scale=1.0]{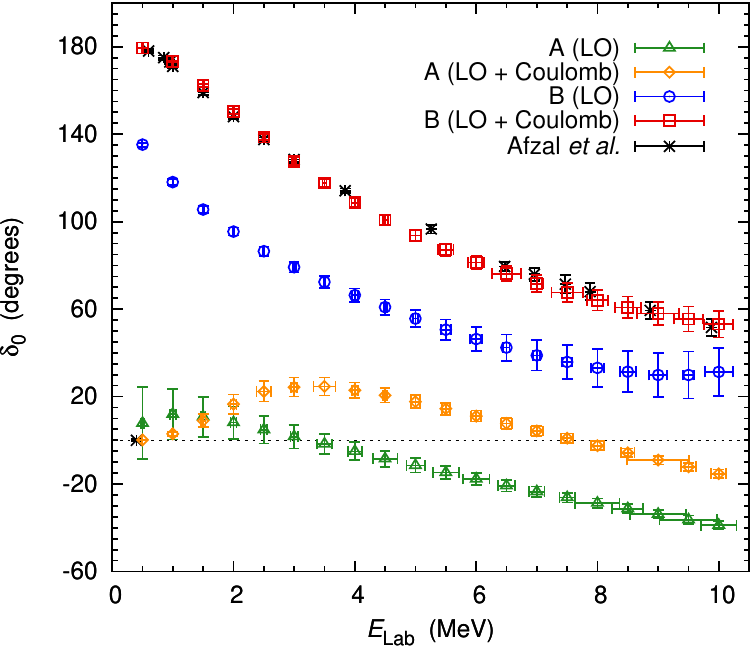}
\label{alpha-alpha}
\end{figure}

To understand how interactions A and  B can produce such completely different
physics, we show their alpha-alpha $S$-wave phase shifts in Fig.~\ref{alpha-alpha}.
 The LO results for interaction A are shown with green triangles, LO + Coulomb
results for A are orange diamonds, LO results for B are blue circles, and
LO + Coulomb results for B are red squares.  The experimental data are shown
with black asterisks \cite{Afzal:1969}.  The phase shifts are computed using
auxiliary-field Monte Carlo simulations and a technique called the adiabatic
projection method \cite{Elhatisari:2015iga}. Interaction B was tuned to
the nucleon-nucleon phase shifts and the alpha-alpha $S$-wave phase shifts,
and so the agreement with experimental data is very good.  However the phase
shifts for interaction A are  small and even negative at larger energies.
 This would explain the large differences between interactions A and B for
the energies of the larger alpha-like nuclei in Table~\ref{nuclei}.

What we have discovered is that alpha-alpha scattering is very sensitive
to the degree of locality of the nucleon-nucleon lattice interactions.  In
Supplemental Materials at [URL will be inserted by publisher] we show that this dependence on degree of locality is due to the
compactness of the alpha-particle wave function.
In contrast, the nucleon-nucleon scattering phase shifts make no constraint
on the degree
of locality of the nucleon-nucleon interactions.  For example, if one starts
with a purely local interaction, a unitary transformation can be used to
define a new interaction which is highly nonlocal but having exactly the
same phase shifts.  The differences only become
apparent in systems with more than two nucleons and can be understood as
arising from three-body and higher-body  interactions \cite{Polyzou:1990a,Bogner:2006pc}.
Interaction A is a perfectly valid starting point for describing nucleon-nucleon
interactions.   However,
substantial higher-nucleon interactions will be needed to rectify the missing
strength of the alpha-alpha interactions and the additional
binding energy in nuclei. 

The results we have found here suggests a strategy for improving future \textit{ab
initio} nuclear structure
and reaction calculations by incorporating low-energy light-nucleus scattering
data in addition to nucleon-nucleon scattering data.
 This would be especially important for accurate calculations of key alpha
capture reactions relevant to astrophysics such as alpha capture on $^{12}$C
\cite{Imbriani:2001xz}.  One can view the extra step of fixing the degree
of locality of the nucleon-nucleon interaction as preemptively reducing the
importance of the required three-body and higher-body interactions.  It 
is similar in spirit to other approaches that
use nuclear structure and many-body observables to help determine the nucleon-nucleon
interactions \cite{Maris:2008ax,Ekstrom:2015rta,Hagen:2015yea}. 

Since alpha-alpha scattering is a difficult and computationally-intensive \textit{ab initio} calculation, it is useful to discuss a simple qualitative picture of the alpha-alpha interaction in a tight-binding approximation. For any nucleon-nucleon interaction $V({\bf r'},{\bf r})$, we define the tight-binding potential, $V_{\rm TB}(r),$ as the contribution that $V({\bf r'},{\bf r})$ makes to the effective interaction between  alpha particles in the tight-binding approximation where the alpha particle radius $R_\alpha$ is treated as a small but non-vanishing length scale.  In this simple approximation the interaction $V({\bf r'},{\bf
r})$ contributes to the  effective alpha-alpha interaction only in two possible ways.  The first is what we call the direct term where $|{\bf r'}-{\bf r}|\lesssim R_{\alpha}$, and the second is the exchange term where $|{\bf
r'}+{\bf r}|\lesssim R_{\alpha}$.  All other terms are forbidden because the interaction is moving the nucleons to locations where there are no alpha particles.  For the LO lattice interactions we consider here at lattice spacing 1.97
fm, we do not attempt to resolve the different microscopic mechanisms that can contribute to $V_{\rm TB}(r)$.  However calculations at smaller lattice spacings would find that the two-pion exchange interaction is responsible for a large attractive tight-binding potential at NNLO \cite{Epelbaum:2008ga}.  This observation connects well
with the work of Ref.~\cite{Arriola:2007de}, which considered the role
of the two-pion exchange interaction in an effective field theory where alpha
particles are treated as fundamental objects. 

\begin{figure}[!h]
\centering
\includegraphics[scale=0.60]{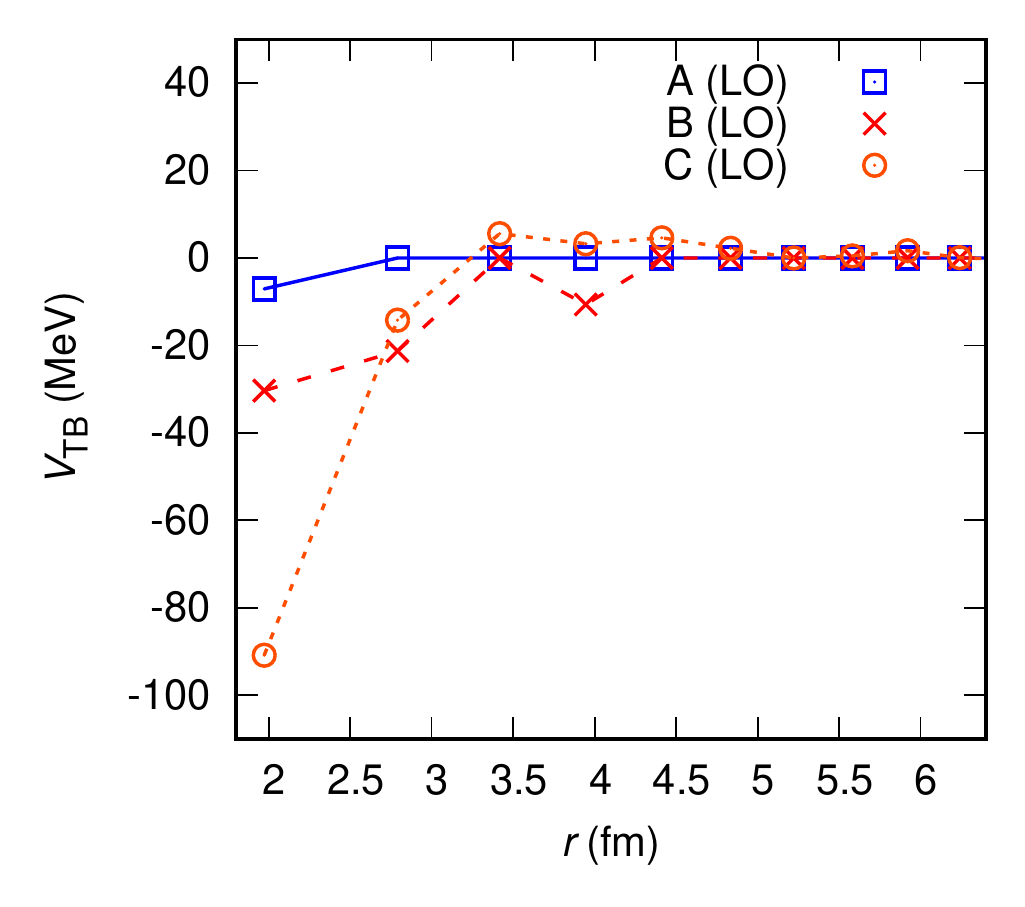}
\caption{Tight-binding potential. We plot the tight-binding
potential versus radial distance for the LO interactions for A, B, and C,
where C is the interaction used in several previous lattice calculations
\cite{Lahde:2013uqa,Elhatisari:2015iga}. Interaction A is shown with blue
squares and solid line, B is drawn with red crosses and dashed line, and
C is presented with orange circles and short-dashed line.}
\label{LTP}
\end{figure}

In Fig.~\ref{LTP} we show the tight-binding potential for the LO lattice interactions for A and B.  For our lattice calculations where space is discrete, we find that $R_\alpha$ is less than one lattice spacing and so the dependence on $R_\alpha$ drops out.  We see that interaction A has a very small tight-binding potential.  This is consistent with the weak alpha-alpha $S$-wave interactions found in Fig.~\ref{alpha-alpha}. In contrast, interaction B has a stronger attractive tight-binding potential resulting from its short-range spin-isospin-independent  local interaction. For comparison we also show in Fig.~\ref{LTP} the tight-binding potential for the leading-order interaction used in prior lattice calculations, which we call interaction C \cite{Lahde:2013uqa,Elhatisari:2015iga}.

In order to discuss the many-body limit, we switch off the Coulomb
interactions and define a one-parameter family of interactions, $V_{\lambda}=(1-\lambda)V_{\rm
A} + \lambda V_{\rm B}$. While the properties of the two, three, and four
nucleon systems vary only slightly with $\lambda$, the many-body ground state
of $V_{\lambda}$ undergoes a quantum phase transition from a Bose-condensed
gas to a nuclear liquid.  

\begin{figure}[!ht]
\centering
\caption{Zero-temperature phase diagram. We show the zero-temperature
phase diagram as a function of the parameter $\lambda$ in the interaction
$V_{\lambda}=(1-\lambda)V_{\rm
A} + \lambda V_{\rm B}$ without Coulomb included.  The blue filled circles
indicate neutrons,
the red filled circles indicate protons, and the small arrows attached to
the circles indicate spin direction.  We show a first-order quantum phase
transition from a Bose gas to nuclear liquid at the point where the scattering
length $a_{\alpha\alpha}$ crosses zero.  We
have also plotted the alpha-like nuclear ground state energies $E_A$ for
$A$ nucleons up to
$A=20$ relative to the corresponding multi-alpha threshold $E_{\alpha}A/4$.
The last
alpha-like nucleus to be bound is $^8$Be at the unitarity point
where $|a_{\alpha\alpha}|=\infty$. \bigskip }
\includegraphics[scale=0.30]{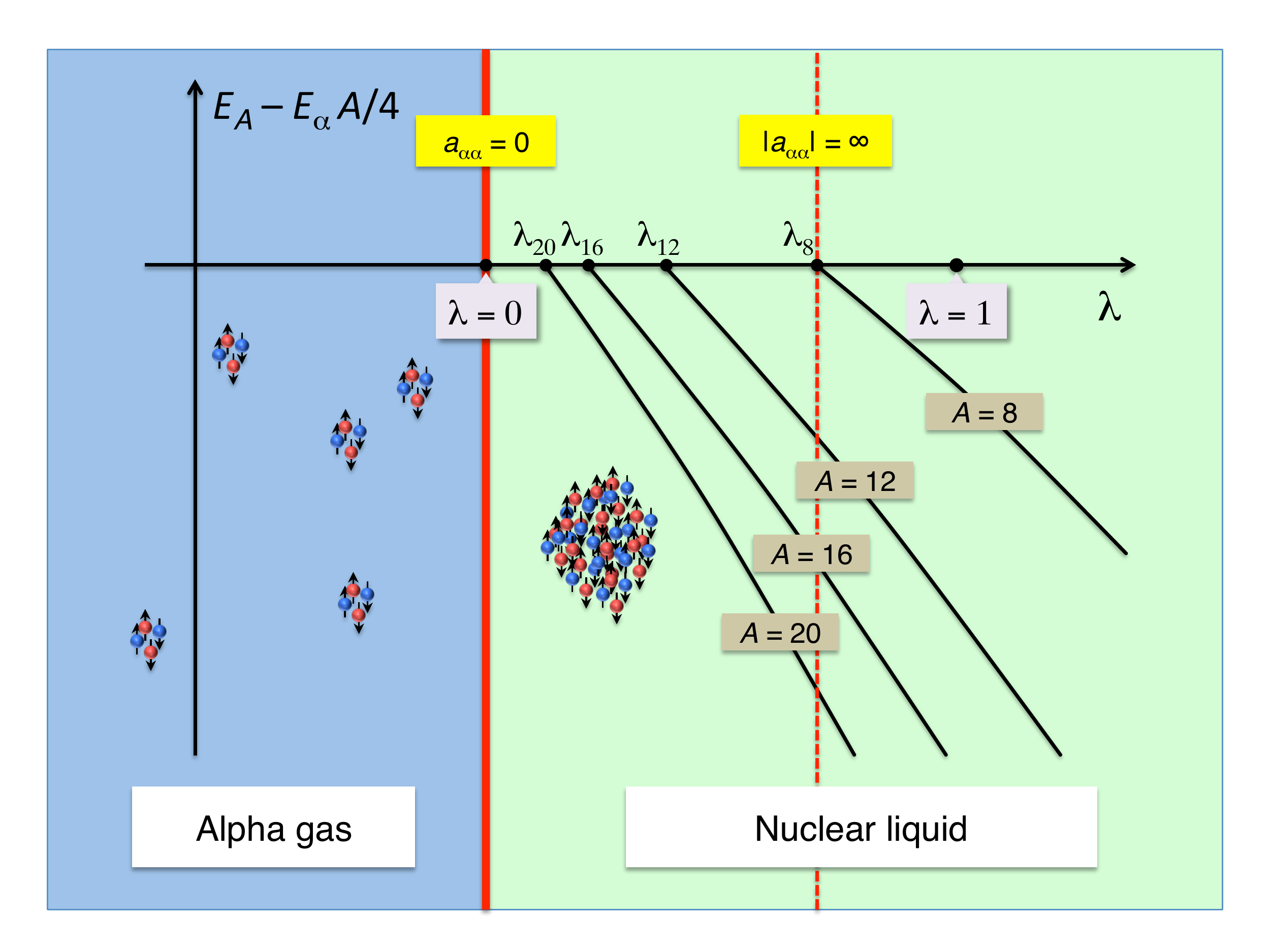}
\label{phase-diagram}
\end{figure}

We sketch the zero temperature phase diagram in Fig.~\ref{phase-diagram}.
The phase transition occurs when the alpha-alpha $S$-wave scattering length
$a_{\alpha\alpha}$
crosses zero, and the Bose gas collapses due to the attractive interactions
\cite{Stoof:1994a,Kagan:1998a}. At slightly larger $\lambda,$ finite alpha-like
nuclei also become bound, starting with the largest nuclei first. The last
alpha-like nucleus to be bound is $^8$Be at the so-called unitarity point
where $|a_{\alpha\alpha}|=\infty$.  Superimposed on the phase diagram, we
have sketched the alpha-like nuclear ground state energies $E_A$ for $A$
nucleons up to $A=20$ relative to the corresponding multi-alpha threshold
$E_{\alpha}A/4$. Empirically we find that the quantum phase transition occurs at the point $\lambda_{\infty} = 0.0(1)$.  The
uncertainty of $\pm 0.1$ is due to the
energy levels having a slow dependence on $\lambda$ near $\lambda = 0.0$. Since any $V_{\lambda}$
represents a seemingly reasonable starting point for the effective field
theory at LO, one may end up crossing the phase transition when considering
higher-order effects beyond LO.  It is in this sense that we
say nature is near a quantum phase transition.

The critical point for the binding of $^{20}$Ne occurs at $\lambda_{20}$~=~$0.2(1)$.
 For the binding of the other alpha nuclei, we obtain  $\lambda_{16}$~=~$0.2(1)$
for $^{16}$O, $\lambda_{12}$~=~$0.3(1)$ for
$^{12}$C, and $\lambda_{8}$~=~$0.7(1)$ for
$^{8}$Be.
One finds a sudden change in the nucleon-nucleon
density correlations at long distances as $\lambda$ crosses the critical point, going from a continuum state to a self-bound system. As $\lambda$ increases further beyond this critical value,
the nucleus becomes more tightly bound, gradually losing its alpha cluster substructure
and becoming more like a nuclear liquid droplet.  The quantum phase transition at $\lambda_{\infty} = 0.0(1)$ is the corresponding phenomenon in the many-body system, a first-order phase transition occurring for infinite matter. 

By adjusting $\lambda$ in  \textit{ab initio} calculations, we have a new
tool for studying alpha cluster states such as the Hoyle state of $^{12}$C
 and possible rotational excitations of the Hoyle state 
 \cite{Funaki:2003af,Chernykh:2007a,Epelbaum:2011md,Zimmerman:2011,Epelbaum:2012qn,Dreyfuss:2012us}.
By tuning $\lambda$ to the unitarity point $|a_{\alpha\alpha}|=\infty$, we
can continuously connect the Hoyle state wave function without Coulomb interactions
to a universal Efimov
trimer \cite{Efimov:1971a,Braaten:2004a,Kraemer:2006Nat}. An Efimov
trimer is one of an infinite tower of three-body states for bosons in the
large scattering-length limit, with intriguing mathematical properties such
as fractal-like discrete scale invariance.  Another interesting system is
 the second $0^+$ state of $^{16}$O \cite{Epelbaum:2013paa}, which should
be continuously connected to a universal Efimov
tetramer \cite{Hammer:2006ct,Kraemer:2006Nat,vonStecher:2009a}.  This connection to Efimov states is now being investigated in follow-up work.  The ability
to tune the energies of alpha cluster states relative to alpha-separation
thresholds provides a new window on wave functions and rotational excitations
of alpha cluster states.  By studying the $\lambda$-dependence of nuclear energy levels one can also identify the underlying cluster substructure.  For example,   the energy of a nuclear state which is a weakly-bound collection of four alpha clusters will track closely with the four-alpha threshold $4E_{^4{\rm He}}$ as a function of $\lambda$, while a state which is comprised of $^{12}$C and $^4$He clusters will track more closely with $E_{^{12}{\rm C}}+E_{^4{\rm He}}$. Such an analysis may provide a new theoretical foundation for understanding clustering in
nuclei and complement existing work on clustering in the literature 
\cite{Wheeler:1937zza,Dennison:1954zz,Clark:1966a,Robson:1979zz,Bauhoff:1984zza,
Tohsaki:2001an}, thereby strengthening the theoretical motivation
for experimental searches of alpha cluster states in alpha-like nuclei. 

\section*{Acknowledgement}
We acknowledge discussions with Hans-Werner Hammer, Thomas Luu, Lucas
Platter, and Thomas Sch{\"a}fer and partial financial support
from the Deutsche Forschungsgemeinschaft (Sino-German CRC 110), the Helmholtz
Association (Contract No.\ VH-VI-417), 
BMBF (Grant No.\ 05P12PDFTE), the U.S. Department of Energy (DE-FG02-03ER41260),
and U.S. National Science Foundation grant No. PHY-1307453.
Further support
was provided by the EU HadronPhysics3 project, the ERC Project No.\ 259218
NUCLEAREFT, the Magnus Ehrnrooth Foundation of the Finnish Society of Sciences
and Letters, MINECO (Spain), the ERDF (European Commission) grant FPA2013-40483-P,
and the Chinese Academy of Sciences (CAS) President's International Fellowship
Initiative (PIFI) grant no. 2015VMA076.
The computational resources 
were provided by the J\"{u}lich Supercomputing Centre at Forschungszentrum
J\"{u}lich and by RWTH Aachen.

\newpage

\section*{Supplemental Materials}

\subsection*{Lattice interactions}
For our LO lattice calculations we use a spatial lattice spacing
$a=(100$~MeV$)^{-1}=1.97$~fm and time lattice step $a_{t}%
=(150$~MeV$)^{-1}=1.32$~fm.  Our axial-vector coupling constant is $g_{A}=1.29$ as derived from the Goldberger-Treiman relation,
the pion decay constant is$\ f_{\pi}=92.2$~MeV, and the pion mass is $M_{\pi}=M_{\pi^{0}}=134.98$~MeV.
\ For the nucleon
mass
we use $m=938.92$ MeV, and the electromagnetic fine structure constant is
$\alpha_{\rm EM}=(137.04)^{-1}$.  We don't consider any isospin-breaking
terms other than the Coulomb interaction in these LO calculations.  We use
$\sigma_S$ with $S=1,2,3$ for the Pauli matrices acting upon spin, and $\tau_I$
with $I=1,2,3$ for the Pauli
matrices acting upon isospin.
We will use lattice units where the quantities are multiplied by the appropriate
power of  the spatial lattice spacing $a$ to make the combination dimensionless.
We write $\alpha_t$ for the ratio $a_t/a$.  We use the notation 
$\sum_{\langle{\bf n'\, n}\rangle}$
to denote the summation over nearest-neighbor lattice sites of {\bf n}.
We write $\sum_{\langle{\bf n'\, n}\rangle_i}$ to indicate the sum over nearest-neighbor
lattice sites of {\bf n} along the $i^{\rm th}$ spatial axis.  Similarly,
we define $\sum_{\langle\langle{\bf n'\, n}\rangle\rangle_i}$ as the sum
over   next-to-nearest-neighbor
lattice sites of {\bf n} along the $i^{\rm th}$ axis and $\sum_{\langle\langle\langle{\bf
n'\, n}\rangle\rangle\rangle_i}$ as the sum
over   next-to-next-to-nearest-neighbor
lattice sites of {\bf n} along the $i^{\rm th}$ axis.
Our lattice geometry is chosen to be an $L^3$ periodic lattice, and so the
summations over ${\bf n'}$ are defined using periodic boundary conditions.

For each lattice site {\bf n} on our lattice and real parameter $s_{\rm NL}$,
we define nonlocal annihilation and creation operators for each spin and
isospin component of the nucleon,
\begin{align}
a_{\rm NL}({\bf n})&=a({\bf n})+s_{\rm NL}\sum_{\langle{\bf n'\, n}\rangle}a({\bf
n'}),\\
a^\dagger_{\rm NL}({\bf n})&=a^\dagger({\bf n})+s_{\rm NL}\sum_{\langle{\bf
n'\, n}\rangle}a^\dagger({\bf
n'}).
\end{align}
For spin indices $S=1,2,3,$ and isospin indices $I=1,2,3$, we define point-like
densities,
\begin{align}
\rho({\bf n})&= a^\dagger({\bf n}) a({\bf n}),
\\
\rho_{S}({\bf n})&=a^\dagger({\bf n})[\sigma_S] a({\bf
n}),
\\
\rho_{I}({\bf n})&=a^\dagger({\bf n})[\tau_I] a({\bf
n}),
\\
\rho_{S,I}({\bf n})&=a^\dagger({\bf n})[\sigma_S \otimes
\tau_I] a_{\rm }({\bf n}).
\end{align}
For spin indices $S=1,2,3,$ and isospin indices $I=1,2,3$, we also define
smeared nonlocal densities, 
\begin{align}
\rho_{\rm NL}({\bf n})&= a^\dagger_{\rm NL}({\bf n}) a_{\rm NL}({\bf n}),
\\
\rho_{S,\rm NL}({\bf n})&=a^\dagger_{\rm NL}({\bf n})[\sigma_S] a_{\rm NL}({\bf
n}),
\\
\rho_{I,\rm NL}({\bf n})&=a^\dagger_{\rm NL}({\bf n})[\tau_I] a_{\rm NL}({\bf
n}),
\\
\rho_{S,I,\rm NL}({\bf n})&=a^\dagger_{\rm NL}({\bf n})[\sigma_S \otimes
\tau_I] a_{\rm NL}({\bf n}),
\end{align}
and smeared local densities for real parameter $s_{\rm L}$, 
\begin{align}
\rho_{\rm L}({\bf n})&= a^\dagger({\bf n}) a({\bf n})+s_{\rm L}\sum_{\langle{\bf
n'\, n}\rangle}a^\dagger({\bf n'}) a({\bf n'}), \\
\rho_{S,\rm L}({\bf n})&=a^\dagger({\bf n})[\sigma_S] a({\bf n})+s_{\rm L}\sum_{\langle{\bf
n'\, n}\rangle}a^\dagger({\bf n'})[\sigma_S] a({\bf n'}),
\\
\rho_{I,\rm L}({\bf n})&=a^\dagger({\bf n})[\tau_I] a({\bf n})+s_{\rm L}\sum_{\langle{\bf
n'\, n}\rangle}a^\dagger({\bf n'})[\tau_I] a({\bf n'}),
\\
\rho_{S,I,\rm L}({\bf n})&=a^\dagger({\bf n})[\sigma_S \otimes \tau_I] a({\bf
n})+s_{\rm L}\sum_{\langle{\bf
n'\, n}\rangle}a^\dagger({\bf n'})[\sigma_S \otimes \tau_I]a({\bf n'}).
\end{align}
The nonlocal short-range interactions are written as
\begin{align}
V_{\rm NL}=\frac{c_{\rm NL}}{2}\sum_{\bf n}:\rho_{\rm NL}({\bf n})\rho_{\rm
NL}({\bf n}):+\frac{c_{I,\rm NL}}{2}\sum_{{\bf n},I}
:\rho_{I,\rm NL}({\bf n})\rho_{I,\rm NL}({\bf n}):,
\end{align}
while the local short-range interactions are
\begin{align}
V_{\rm L}= & \frac{c_{\rm L}}{2}\sum_{\bf n}:\rho_{\rm L}({\bf n})\rho_{\rm
L}({\bf n}):+\frac{c_{S,\rm L}}{2}\sum_{{\bf n},S} :\rho_{S,\rm L}({\bf n})\rho_{S,\rm
L}({\bf n}): \nonumber \\
& +\frac{c_{I,\rm L}}{2}\sum_{{\bf n},I}
:\rho_{I,\rm L}({\bf n})\rho_{I,\rm L}({\bf n}):+\frac{c_{S,I,\rm L}}{2}\sum_{{\bf
n},S,I}
:\rho_{S,I,\rm L}({\bf n})\rho_{S,I,\rm L}({\bf n}):.
\end{align}
The $::$ symbol indicates normal ordering, where the annihilation operators
are on the right-hand side and the creation operators are on the left-hand
side. As described in previous work \cite{Epelbaum:2010xt}, we take special
combinations of the four local short-range operator coefficients so that
the interaction in odd partial waves vanish completely.  For our work here,
we also make the strength of the local short-range interactions equal in
the two $S$-wave channels.  As a result, we have only one independent coefficient,
$c_{S,{\rm L}}=c_{I,{\rm L}}=c_{S,I,{\rm L}}=-\frac{1}{3} c_{\rm L}.$  In
future work it may be useful to consider relaxing this condition.   

The one-pion exchange interaction has the form
\begin{align}
V_{\rm OPE}=-\frac{g_A^2}{8f^2_{\pi}}\sum_{{\bf n',n},S',S,I}
:\rho_{S',I\rm }({\bf n'})f_{S'S}({\bf n'}-{\bf n})\rho_{S,I}({\bf n}):,
\end{align}
where $f_{S'S}$ is defined as
\begin{align}
f_{S'S}({\bf n'}{\bf -n})=\frac{1}{L^3}\sum_{\bf q}\frac{\exp[-i{\bf q}\cdot({\bf
n'}-{\bf n})-b_{\pi}{\bf q}^2]q_{S'}q_{S}}{{\bf q}^2 + M_{\pi}^2},
\end{align}
and each lattice momentum component $q_S$ is an integer multiplied by $2\pi/L$.
 The parameter $b_{\pi}$ is included to remove short-distance lattice artifacts
in the one-pion exchange interaction.  It results in better preservation
of rotational symmetry and will be especially useful at smaller lattice spacings
\cite{Klein:2015vna}.  The Coulomb interaction can be written as
\begin{align}
V_{\rm Coulomb}=-\frac{\alpha_{\rm EM}}{2_{}}\sum_{{\bf n',n}}
:\frac{1}{4}[\rho({\bf n'})+\rho_{I=3\rm }({\bf n'})]\frac{1}{d({\bf
n'}-{\bf n})}[\rho({\bf n})+\rho_{I=3\rm }({\bf n})]:,
\end{align}
where $d({\bf
n'}-{\bf n})$ is the shortest length of ${\bf
n'}-{\bf n}$ as measured on the periodic lattice, and we define the value
of $d$ at the
origin to be $\frac{1}{2}$.
Our notation $\rho_{I=3\rm }$ refers to the $I=3$ isospin component of $\rho_{I}$.
We use a free lattice Hamiltonian \cite{Epelbaum:2010xt}
of the form,
\begin{align}
H_{\rm free}= &\frac{49}{12m}\sum_{\bf n} a^\dagger({\bf n}) a({\bf n})-\frac{3}{4m}\sum_{{\bf
n},i}
\sum_{\langle{\bf n'}\,{\bf n}\rangle_i} a^\dagger({\bf n'}) a({\bf n}) \nonumber \\
&+\frac{3}{40m}\sum_{{\bf
n},i}\sum_{\langle\langle{\bf n'}\,{\bf n}\rangle\rangle_i} a^\dagger({\bf
n'})
a({\bf n})-\frac{1}{180m}\sum_{{\bf
n},i}
\sum_{\langle\langle\langle{\bf n'}\,{\bf n}\rangle\rangle\rangle_i} a^\dagger({\bf
n'}) a({\bf n}).
\end{align}
For interaction A at LO, the lattice Hamiltonian is
\begin{align}
H_{\rm A} = H_{\rm free} + V_{\rm NL} + V_{\rm OPE},
\end{align}
with $s_{\rm NL}=0.07700$, $c_{\rm NL}=-0.2268$, $c_{I,\rm NL}=0.02184$,
and $b_{\pi}=0.7000$.
These parameters are determined by fitting to the low-energy nucleon-nucleon
phase shifts and the observed deuteron energy. For the corresponding LO +\
Coulomb interactions,
we
simply add $V_{\rm Coulomb}$ to $H_{\rm A}$.
 
For interaction B at LO, we have
\begin{align}
H_{\rm B} = H_{\rm free} + V_{\rm NL} + V_{\rm L} + V_{\rm OPE},
\end{align}
with $s_{\rm NL}=0.07700$, $s_{\rm L}=0.8100$, $c_{\rm NL}=-0.1171$, $c_{I,\rm
NL}=0.02607$, $c_{\rm L}=-0.01013$, and $b_{\pi}=0.7000$.  For the corresponding
LO +\ Coulomb interactions,
we
simply add $V_{\rm Coulomb}$ to $H_{\rm B}$. These parameters are determined
by fitting to the low-energy nucleon-nucleon
phase shifts, the observed deuteron energy, and the low-energy alpha-alpha
$S$-wave phase shifts. 

We should clarify that the $^4$He energy is not used to fit the parameters
of interactions A and B. However we do observe a strong correlation between
the alpha-alpha $S$-wave phase shifts and the shape of the $^4$He wave function
tail.  This has the resulting effect of driving the $^4$He energy close to
the physical value when we tune the parameters of interaction B to the alpha-alpha
$S$-wave phase shifts.  The parameters of interaction A are determined by
starting from the parameters of interaction B,  setting the local short-range
interactions to zero, and then tuning the coefficients of the nonlocal short-range
interactions to the nucleon-nucleon phase shifts and deuteron energy.

\subsection*{Nucleon-nucleon scattering}

We use the spherical wall method to calculate lattice phase shifts \cite{Carlson:1984,Borasoy:2007vy}.
 We use the  improvements recently introduced in Ref.~\cite{Lu:2015riz}.
Let $|{\bf n}\rangle
\otimes |S_z \rangle$ be the two-nucleon scattering state with lattice separation
vector
${\bf n}$ and $z$-component of total intrinsic spin $S_z$. We define radial
coordinates on the lattice by grouping together
lattice mesh points with the same radial distance to define radial position
states and project onto states with total angular momentum $J,J_z$
in the continuum limit.  Using spherical harmonics
$Y_{\ell,\ell_z}$ with orbital angular momentum $\ell,\ell_z$ and Clebsch-Gordan
coefficients
$C^{J,J_z}_{\ell,\ell_z,S,S_z}$,
we define \begin{equation}
        |r\rangle^{J,J_z}_L = \sum_{{\bf n},\ell_z,S_z}C^{J,J_z}_{\ell,\ell_z,S,S_z}Y_{\ell,\ell_z}^{\,}(\hat{{\bf
n}})\delta_{r,|{\bf n}|}|{\bf n}\rangle
\otimes |S_z \rangle, \label{spherical}
\end{equation}
where $\delta_{r,|{\bf n}|}$
is a Kronecker delta function that selects lattice points where $|{\bf n}|
= r$. This angular momentum projection allows us to calculate partial-wave
phase shifts on the lattice as in Ref.~\cite{Lu:2015riz}.  

As described in Ref.~\cite{Lu:2015riz}, we impose a hard spherical
wall boundary at some large radius $R_W$  and a smooth auxiliary Gaussian
potential in front of the wall, which we call $V_{\rm aux}(r)$.  For our
calculations here we use $R_W=15.02$ lattice units.  The auxiliary potential
has the form 
\begin{align}
V_{\rm aux}(r)=V_0 \exp\left[-(r-R_W)^2\right],
\end{align}
with adjustable~coefficient $V_0$ that is used to probe different values
of the scattering energy.  The auxiliary potential is non-negligible only
when $r$ is a few lattice units away the wall at $R_W$.  We determine the
asymptotic phase shifts from the radial wave function at points where $r$
is large but $V_{\rm aux}(r)$ is negligible.
For coupled partial waves such as the ${^3s_1}-{^3d_1}$ channel, we determine
the two phase shifts and mixing angle  using an additional auxiliary potential
$U_{\rm aux}(r)$ with the same functional form  as $V_{\rm aux}(r),$ but
with imaginary Hermitian off-diagonal couplings between the two partial waves,
\begin{align}
\begin{bmatrix}
    0       & iU_{\rm aux}(r) \\
     -iU_{\rm aux}(r)   &  0\\
\end{bmatrix}.
\end{align}
This complex-valued 
auxiliary potential breaks time-reversal invariance and allows us to extract
information about the two independent phase shifts and mixing angle from
the real and imaginary parts of the complex-valued wave functions. 

\begin{figure}[!ht]
\centering
\caption{Nucleon-nucleon scattering phase shifts. We plot LO lattice
phase shifts for proton-neutron scattering versus the center-of-mass relative
momentum for interactions A (red triangles) and B (blue squares).  For comparison
we also plot the phase shifts extracted from the Nijmegen partial wave analysis
\cite{Stoks:1993tb} (black lines) and a continuum version of interaction
A (green dashed lines).  In the first row, the data in panels {\bf a}, {\bf
b}, {\bf c}, {\bf d} correspond to ${^1S_0},{^3S_1}, {^1P_1}, {^3P_0}$ respectively.
In the second row, panels {\bf e}, {\bf f}, {\bf g}, {\bf h} correspond to
${^3P_1},{^3P_2},
{^1D_2}, {^3D_1}$ respectively.   In the third row, panels {\bf i}, {\bf
j}, {\bf k}, {\bf l} correspond to ${^3D_2},{^3D_3},
{\varepsilon_1}, {\varepsilon_2}$ respectively. \bigskip}
\includegraphics[scale=1.5]{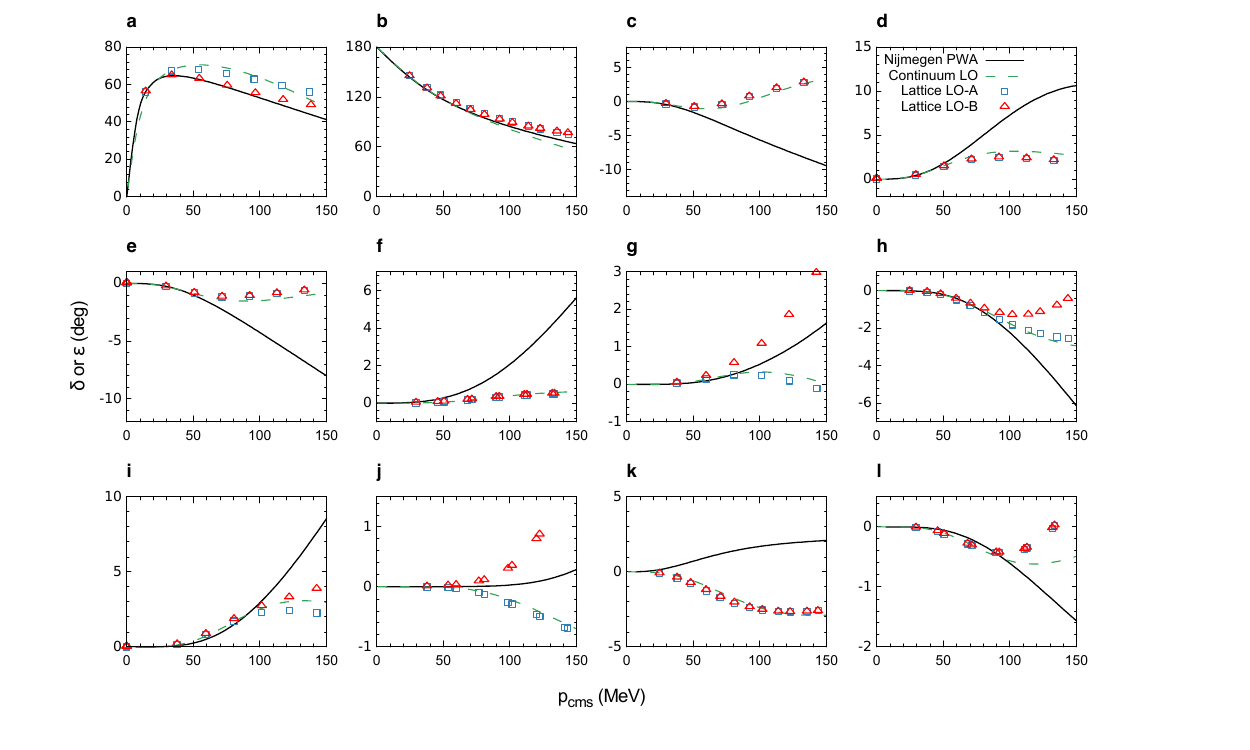}
\label{NN}
\end{figure}

In Fig.~\ref{NN} we show LO lattice
phase shifts for proton-neutron scattering versus the center-of-mass relative
momentum for interactions A (red triangles) and B (blue squares).
 For comparison we also plot the phase shifts extracted from the Nijmegen
partial wave analysis \cite{Stoks:1993tb} (black lines) and a continuum
version of interaction A (green dashed lines).  In the first row, the
data in panels {\bf a}, {\bf b}, {\bf c}, {\bf d} correspond to ${^1S_0},{^3S_1},
{^1P_1}, {^3P_0}$ respectively. In the second row, panels {\bf e}, {\bf f},
{\bf g}, {\bf h} correspond to ${^3P_1},{^3P_2},
{^1D_2}, {^3D_1}$ respectively.   In the third row, panels {\bf i}, {\bf
j}, {\bf k}, {\bf l} correspond to ${^3D_2},{^3D_3},
{\varepsilon_1}, {\varepsilon_2}$ respectively.  The level of agreement with
the experimental phase shifts for interactions A and B is typical
for LO chiral effective field theory at our cutoff momentum of $\pi/a \approx 314\;{\rm MeV}$.  The agreement would be somewhat better if we were to use a smaller value of the smearing parameter $b_{\pi}$ in the one-pion exchange potential.  However, we prefer the higher value of $b_{\pi}$ to reduce sign oscillations   in the Monte Carlo lattice simulations.  The LO interactions  are more than sufficient
to illustrate the ideas of this work but not sufficient for precision calculations.
 For precision calculations, this would be just the first step in the chiral
effective field theory expansion, and the phase shifts would be systematically
improved at each higher order, NLO, NNLO, and so on.

We note the good agreement between the continuum results in green dashed
lines and lattice interaction A results.  This is a good indication that
we have successfully reduced lattice artifacts from the calculations and
was part of the motivation for introducing the parameter $b_{\pi}$.  The
nonlocal smeared interaction $V_{\rm NL}$ makes a non-negligible contribution
to the $S$-wave interactions only. Furthermore, the local smeared interaction
$V_{\rm L}$ makes a nonzero contribution to only the even partial waves ($S$,
$D$, $\cdots$).  Hence the interactions A and B are exactly the same in all
odd partial waves.  We see that the $S$-wave interactions for interactions
A and B are also quite similar, though the ${^1S_0}$ partial wave scattering
is somewhat more attractive for interaction A.  On the other hand, the $D$-wave
partial waves are more attractive for interaction B.  
\subsection*{Euclidean time projection and auxiliary-field Monte Carlo} 
In these lattice simulations we work with the Euclidean time transfer matrix
$M$, which is defined as the normal-ordered exponential of the lattice Hamiltonian
$H$ over one time lattice step,
\begin{align}
M = :\exp[-H\alpha_t]:.
\end{align}
We consider some initial state $|\Psi_i\rangle $ and final state $|\Psi_f\rangle$
 that have nonzero overlap with the ground state of interest.  By applying
successive powers of $M$ upon $|\Psi_i\rangle$, the excited states decay
away, and we can project out only the ground state.  We calculate projection
amplitudes of the form
\begin{equation}
A_{fi}(L_t) = \langle \Psi_f| M^{L_t}  |\Psi_i\rangle.
\end{equation}
By calculating the ratio $A_{fi}(L_t)/A_{fi}(L_t-1)$ for large $L_t$ we can
determine the ground state energy.  In order to calculate first-order corrections
to the ground state energy due to an additional term $\Delta H$ in the Hamiltonian,
we also calculate the projection amplitude 
\begin{align}
A^{\Delta}_{fi}(L_t) = \langle \Psi_f| M^{\frac{L_t-1}{2}} M_{\Delta} M^{\frac{L_t-1}{2}}
 |\Psi_i\rangle,
\end{align}
for odd $L_t$, where\begin{align}
M_{\Delta} &= :\exp[-(H+\Delta H)\alpha_t]:.
\end{align}
The
corrections due to $H_{\rm Coulomb}$ are computed in this manner.

In most cases it is advantageous to first prepare
the initial state using a simpler transfer matrix $M_*$ which
is an approximation to $M$.  We choose $M_*$ to be invariant under Wigner's
SU$(4)$ symmetry \cite{Wigner:1937} where the four spin-isospin combinations
of the nucleon transform into one another.  The SU(4) symmetry eliminates
sign oscillations  from auxiliary-field Monte Carlo simulations of $M_*$ \cite{Chen:2004rq,Lee:2007eu}.    
$M_*$ has the same form as $M,$ but the coefficients of operators that violate
SU(4) symmetry are turned off.  We use $M_*$ as an approximate low-energy
filter by multiplying the initial and final states by $M_*$ for some fixed
number of times, $L_t'$,
  \begin{equation}
A_{fi}(L_t) = \langle \Psi _f| M_*^{L'_t}   M^{L_t}   M_*^{L'_t}  |\Psi_i\rangle.
\label{mstar2}\end{equation} 

We use auxiliary fields to generate the interactions contained in our lattice
Hamiltonian. \ The auxiliary field method can be understood as a Gaussian
integral formula which relates the exponential
of the two-particle density, $\rho^2$, to the integral of the exponential
of the one-particle density, $\rho$,\begin{equation}
{:\exp\left(-\frac{c\alpha_t}{2}\rho^2\right):}=\sqrt{\frac{1}{2\pi}}\int^{\infty}_{-\infty}ds
\, {:\exp \left(-\frac{1}{2}s^2 + \sqrt{-c\alpha_t}s\rho \right):}\;. \label{HS}
\end{equation}
The normal ordering symbol :: ensures that the operator products of the creation
and annihilation operators are treated as though classical anticommuting
Grassmann variables~\cite{Lee:2008fa}. We use this integral identity to
introduce auxiliary fields defined over every lattice point in space and
time \cite{Hubbard:1959ub,Stratonovich:1958,Koonin:1986}. As we will see
shortly, the pion fields are treated in a manner similar to the auxiliary
fields. Each nucleon
is independent of the other nucleons and  interacts only with the auxiliary
and pion
fields. If the initial and final states are an antisymmetric tensor product
of $A$ single nucleon states, then the projection amplitude for any configuration
of auxiliary and pion fields is
proportional to the determinant of an $A\times A$ matrix $Z_{jk}$. \ The
matrix entries of $Z_{jk}$ are single nucleon amplitudes for a nucleon
starting at state $k$ and ending at state $j$.
 
We couple auxiliary fields $s$ to $\rho_{\rm NL}$ and $s_I$ to $\rho_{I,\rm
NL}$ for the nonlocal interactions in $V_{\rm NL}$.  The terms linear in
the auxiliary fields are
\begin{equation}
V^s_{\rm NL} = \sqrt{-c_{\rm NL}} \sum_{{\bf n}}\rho_{\rm NL}({\bf n})s({\bf
n})+\sqrt{-c_{I,\rm NL}} \sum_{{\bf n},I}\rho_{I,\rm NL}({\bf
n})s_I({\bf n}),
\end{equation}
and the terms quadratic field in the auxiliary fields are
\begin{equation}
V^{ss}_{\rm NL} = \frac{1}{2} \sum_{{\bf n}}s^2({\bf
n})+\frac{1}{2} \sum_{{\bf n},I}s^2_I({\bf n}). \\ 
\end{equation}
We also couple auxiliary fields $u$ to $\rho_{\rm L}$, $u_S$ to $\rho_{S,\rm
L}$, $u_I$ to $\rho_{I,\rm
L}$, and $u_{S,I}$ to $\rho_{S,I,\rm
L}$, for the local interactions in $V_{\rm L}$, 
\begin{align}
V^u_{\rm L} = & \sqrt{-c_{\rm L}} \sum_{{\bf n}}\rho_{\rm L}({\bf n})u({\bf
n})+\sqrt{-c_{S,\rm L}} \sum_{{\bf n},S}\rho_{S,\rm L}({\bf
n})u_S({\bf n}) \nonumber \\
& +\sqrt{-c_{I,\rm L}} \sum_{{\bf n},I}\rho_{I,\rm L}({\bf
n})u_I({\bf n})+\sqrt{-c_{S,I,\rm L}} \sum_{{\bf n},S,I}\rho_{S,I,\rm L}({\bf
n})u_{S,I}({\bf n}),
\end{align}
\begin{equation}
V^{uu}_{\rm L} = \frac{1}{2} \sum_{{\bf n}}u^2({\bf
n})+\frac{1}{2} \sum_{{\bf n},S}u^2_S({\bf n})+\frac{1}{2} \sum_{{\bf n},I}u^2_I({\bf
n})+\frac{1}{2} \sum_{{\bf n},{S,I}}u^2_{S,I}({\bf n}). \\ 
\end{equation}
For the one-pion exchange interaction we couple the gradient of the pion
field $\pi_I$ to
the point-like density $\rho_{S,I}$, 
\begin{equation}
V^{\pi}=\frac{g_A}{2f_{\pi}} \sum_{{\bf n},S,I}\rho_{S,I}({\bf
n'})f^{\pi}_{S}({\bf n'-n)}\pi_{I}({\bf n}),
\end{equation}
\begin{equation}
V^{\pi\pi}=\frac{1}{2}{} \sum_{{\bf n},I}\pi_I({\bf
n'})f^{\pi\pi}_{}({\bf n'-n)}\pi_{I}({\bf n}):,
\end{equation}
where $f^{\pi}_{S}$ and $f^{\pi \pi}$ are defined as
\begin{equation}
f^{\pi}_{S}({\bf n'}{\bf -n})=\frac{1}{L^3}\sum_{\bf q}\exp[-i{\bf
q}\cdot({\bf
n'}-{\bf n})]q_S,
\end{equation} 
\begin{equation}
f^{\pi\pi}({\bf n'}{\bf -n})=\frac{1}{L^3}\sum_{\bf q}\exp[-i{\bf q}\cdot({\bf
n'}-{\bf n})+b_{\pi}{\bf q}^2]({\bf q}^2 + m_{\pi}^2).
\end{equation}Then the transfer matrices for the LO\ interactions can be
written in following manner.  For interaction A we have 
\begin{align}
:\exp\left(-H_A\alpha_t\right):=\int DsD\pi:\exp\left(-H_{\rm free}\alpha_t-V^s_{\rm
NL}\sqrt{\alpha_t}-V^{ss}_{\rm
NL}-V^{\pi}\alpha_t-V^{\pi\pi}\alpha_t\right):,
\end{align}
where $Ds$ is the path integral measure for $s$ and $s_I$, and $D\pi$ is
the path integral measure for $\pi_I$. For interaction B we find
\begin{align}
& :\exp\left(-H_B\alpha_t\right):= \nonumber \\ 
& \int DsDuD\pi:\exp\left(-H_{\rm free}\alpha_t-V^s_{\rm
NL}\sqrt{\alpha_t}-V^{ss}_{\rm
NL}-V^u_{\rm
L}\sqrt{\alpha_t}-V^{uu}_{\rm
L}-V^{\pi}\alpha_t-V^{\pi\pi}\alpha_t\right):,
\end{align}
where $Du$ is the path integral measure for $u$, $u_S$, $u_I$, and $u_{S,I}$.
See Ref.~\cite{Lee:2008fa} for details on the Monte Carlo importance sampling
algorithms used to calculate the path integrals over the auxiliary and pion
fields.

When computing the energy from ratios of amplitudes $A_{fi}(L_t)/A_{fi}(L_t-1)$,
previous studies have used importance sampling according to the importance
function $|A_{fi}(L_t-1)|$ or $|A_{fi}(L_t)|$.  In this work we sample according
to a linear combination $x|A_{fi}(L_t-1)|+(1-x)|A_{fi}(L_t)|$ where $0<x<1$.
 This greatly reduces the stochastic noise because the contributions to $A_{fi}(L_t-1)$
and $A_{fi}(L_t)$ from any individual configuration are now bounded above
in magnitude,
\begin{align}
\frac{|A_{fi}(L_t-1)|}{x|A_{fi}(L_t-1)|+(1-x)|A_{fi}(L_t)|}&<x^{-1}, \\
\frac{|A_{fi}(L_t)|}{x|A_{fi}(L_t-1)|+(1-x)|A_{fi}(L_t)|}&<(1-x)^{-1}.
\end{align}
 
\subsection*{Ground state energies of nuclei}

\begin{figure}[!ht]
\centering
\caption{Energy versus projection time for ${^3}$H,${^3}$He, and ${^4}$He.
In panels {\bf a} and {\bf b} we plot the
energy versus projection time $t = L_ta_t$ for $^{3}$He and $^{4}$He respectively
for the LO interaction A (blue plus signs and dashed lines), LO interaction
B (red squares
and dashed lines), LO + Coulomb interaction A (blue crosses and solid lines),
and  LO + Coulomb
interaction B (red triangles and solid lines).  The LO and LO + Coulomb results
${^3}$H are the same as
the LO results for ${^3}$He. The error bars indicate one standard deviation
errors from the stochastic noise of the Monte Carlo simulations, and
the lines show extrapolations to infinite projection time. \bigskip}
\includegraphics[scale=0.7]{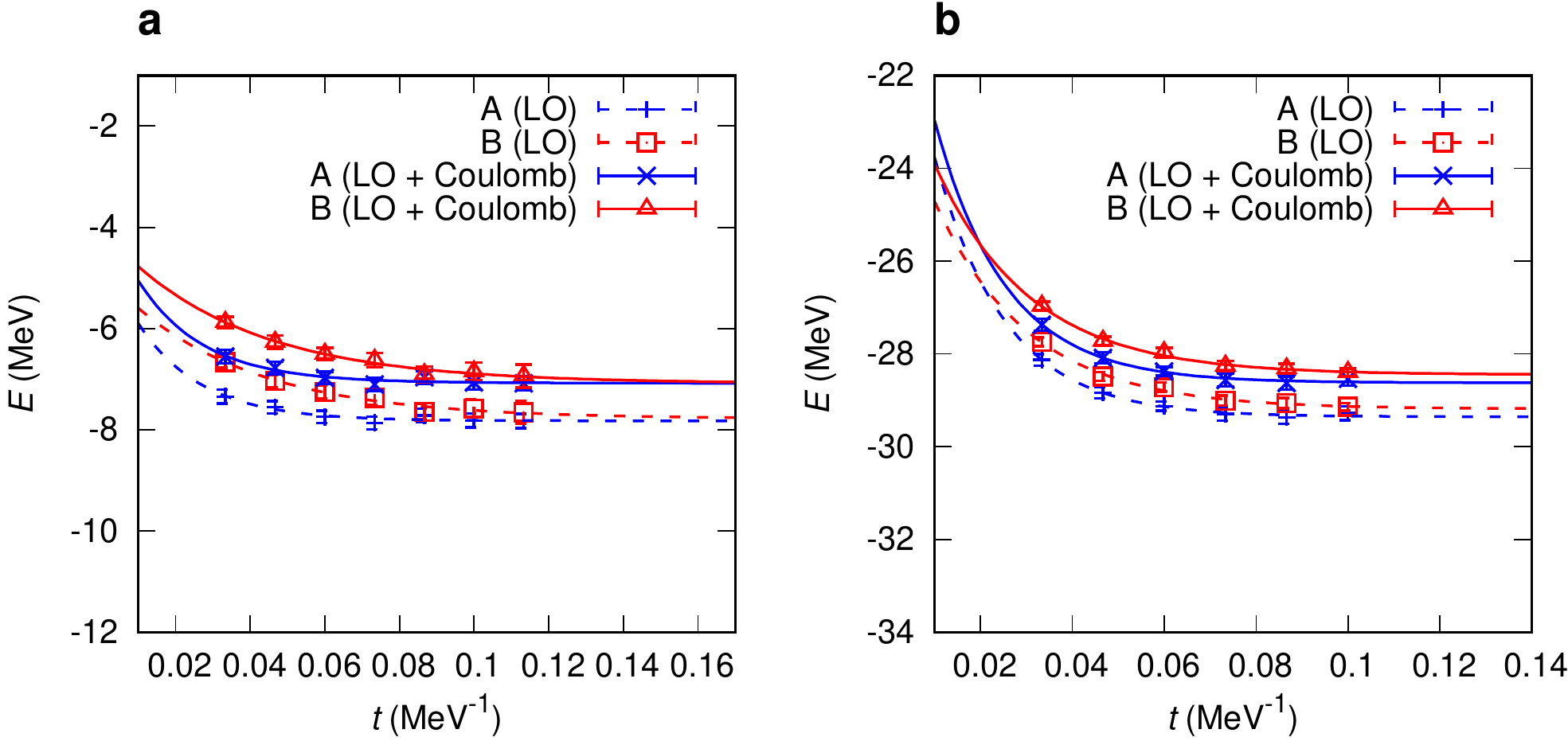}
\label{He3_He4}
\end{figure}

We let $a^\dagger_{\uparrow,p}({\bf n})$, 
$a^\dagger_{\downarrow,p}({\bf n}),$ 
$a^\dagger_{\uparrow,n}({\bf n})$, and $a^\dagger_{\downarrow,n}({\bf n})$
be the  creation operators for a spin-up proton, spin-down proton, spin-up
neutron, and spin-down neutron.  We write $\tilde{a}^\dagger_{\uparrow,p}(0),$
$\tilde{a}^\dagger_{\downarrow,p}(0),$ $\tilde{a}^\dagger_{\uparrow,n}(0)$,
and $\tilde{a}^\dagger_{\downarrow,n}(0)$
for the corresponding zero-momentum creation operators. We also write $\prod\tilde{a}^\dagger$
as shorthand for the product
\begin{equation}
\prod\tilde{a}^\dagger=\tilde{a}^\dagger_{\uparrow,p}(0)\tilde{a}^\dagger_{\downarrow,p}(0),
\tilde{a}^\dagger_{\uparrow,n}(0)\tilde{a}^\dagger_{\downarrow,n}(0).
\end{equation}
For the ground state energy calculations of $^3$H and $^3$He we use a lattice
volume of $(16\;{\rm fm})^3$.  The initial states we choose are
\begin{align}
|\Psi^{{^3}{\rm H}}_i\rangle &=\sum_{\bf n,n',n'',n'''}
e^{-\alpha|{\bf n-\bf n'|}}
e^{-\alpha|{\bf n-\bf n''|}}
e^{-\alpha|{\bf n-\bf n'''|}}
a^\dagger_{\uparrow,p}
({\bf n'})
a^\dagger_{\uparrow,n}({\bf n''})a^\dagger_{\downarrow,n}({\bf n'''})\left|0\right>,
\\
|\Psi^{{^3}{\rm He}}_i\rangle &=\sum_{\bf n,n',n'',n'''}
e^{-\alpha|{\bf n-\bf n'|}}
e^{-\alpha|{\bf n-\bf n''|}}
e^{-\alpha|{\bf n-\bf n'''|}}
a^\dagger_{\uparrow,n}({\bf n'})
a^\dagger_{\uparrow,p}({\bf n''})a^\dagger_{\downarrow,p}({\bf n'''})\left|0\right>,
\end{align}
with $\alpha = 2$ in lattice units.
In panel {\bf a} of Fig.~\ref{He3_He4}
we show the energy versus projection time $t = L_ta_t$ for ${^3}$He for the
LO interaction A (blue plus signs and dashed lines), LO interaction B (red
squares and dashed lines), LO + Coulomb interaction A (blue crosses and solid
lines), and  LO + Coulomb
interaction B (red triangles and solid lines).  As we are not including isospin-breaking
effects other than Coulomb interactions, the LO and LO + Coulomb results
for ${^3}$H are exactly the same as
the LO results for ${^3}$He. The error bars indicate one standard deviation
errors due to the stochastic noise of the Monte Carlo simulations. The lines
are extrapolations to infinite projection time using the ansatz,
\begin{equation}
E(t) = E_{0} + c\exp[-\Delta E \, t],
\label{E_extrap}
\end{equation}
where $E_{0}$ is the ground state energy that we wish to determine.  The
results for the ground state energies are shown in Table~1.

For the ground state energy calculations of $^4$He we use a lattice
volume of $(12\;{\rm fm})^3$.  The initial state we choose is
\begin{align}
|\Psi^{{^4}{\rm He}}_i\rangle &=\prod\tilde{a}^\dagger\left|0\right>.
\end{align}In panel {\bf b} of Fig.~\ref{He3_He4}
we show the energy versus projection time $t = L_ta_t$ for ${^4}$He for the
LO
interaction A (blue plus signs and dashed lines), LO interaction B (red squares
and dashed lines), LO + Coulomb interaction A (blue crosses and solid lines),
and  LO + Coulomb
interaction B (red triangles and solid lines).  The error bars indicate one
standard deviation errors of the Monte Carlo simulations, and the lines are
extrapolations to infinite projection time using
the ansatz in Eq.~(\ref{E_extrap}).  The results for the ground state energies
are shown in Table~1.

We note that while that the $^3$H energies for interactions A and B are underbound,
the energies for $^4$He are near the physical value. This may seem puzzling
since in continuum-space calculations there is a well-known universal correlation
between the $^3$H and $^4$He energies called the Tjon line \cite{Tjon:1975a,Nogga:2000uu,Platter:2004zs}.
 Our lattice results show some deviation from this universal behavior due
to lattice artifacts associated with our lattice spacing of 1.97 fm.  This
is not a new observation.   The same behavior has been analyzed previously
at the same lattice spacing but with a different lattice interaction \cite{Epelbaum:2009pd,Epelbaum:2010xt}.
In order to match the physical $^3$H and $^4$He energies  at the same time,
higher-order short-range three-nucleon interactions at N$^4$LO and possibly
the leading-order short-range four-nucleon interaction at N$^5$LO are needed.
 However a much simpler solution is to use a smaller lattice spacing, as
these lattice deviations from the continuum-space Tjon line decrease very
rapidly with the lattice spacing.

\begin{figure}[!ht]
\centering
\caption{Energy versus projection time for ${^8}$Be,$^{12}$C,$^{16}$O,
and $^{20}$Ne. In panels {\bf a}, {\bf b}, {\bf c}, {\bf d} we plot the
energy versus projection time $t = L_ta_t$ for $^{8}$Be, $^{12}$C,
$^{16}$O, and $^{20}$Ne respectively for the
LO
interaction A (blue plus signs and dashed lines), LO interaction B (red squares
and dashed lines), LO + Coulomb interaction A (blue crosses and solid lines),
and  LO + Coulomb
interaction B (red triangles and solid lines).  The error bars indicate one
standard deviation errors from the stochastic noise of the Monte Carlo simulations,
and
the lines show extrapolations to infinite projection time. \bigskip } 
\includegraphics[scale=0.7]{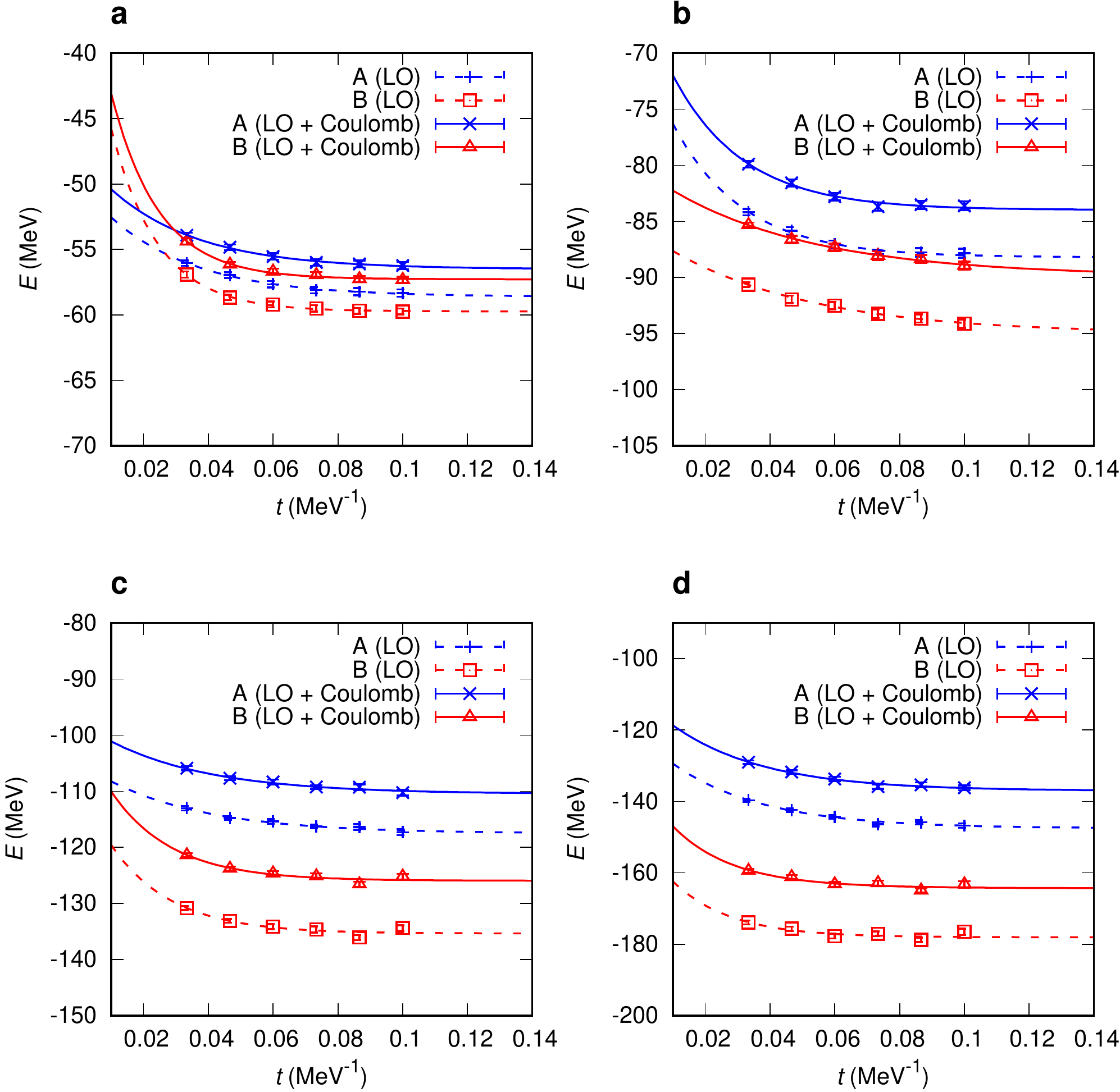}
\label{Be8_C12_O16_Ne20}
\end{figure}

For the ground state energy calculations of $^8$Be, $^{12}$C, $^{16}$O, and
$^{20}$Ne we use a lattice
volume of $(12\;{\rm fm})^3$.  The initial states we use are
\begin{align}
|\Psi^{{^8}{\rm Be}}_i\rangle &=\prod\tilde{a}^\dagger \cdot M_*\prod\tilde{a}^\dagger\left|0\right>,
\\
|\Psi^{{^{12}}{\rm C}}_i\rangle &= \prod\tilde{a}^\dagger \cdot M_*\prod\tilde{a}^\dagger
\cdot M_* \prod\tilde{a}^\dagger\left|0\right>,\\
|\Psi^{{^{16}}{\rm O}}_i\rangle &=\prod\tilde{a}^\dagger \cdot M_*\prod\tilde{a}^\dagger
\cdot M_*\prod\tilde{a}^\dagger \cdot M_*\prod\tilde{a}^\dagger \left|0\right>,\\
|\Psi^{{^{20}}{\rm Ne}}_i\rangle &=\prod\tilde{a}^\dagger \cdot M_*\prod\tilde{a}^\dagger
\cdot M_*\prod\tilde{a}^\dagger \cdot M_*\prod\tilde{a}^\dagger \cdot M_*\prod\tilde{a}^\dagger
\left|0\right>.
\end{align}
The interspersing of the transfer matrix $M_*$ in between the products of
creation operators allows us to create all nucleons with zero momentum without
violating the Pauli exclusion principle.
In panels {\bf a, b, c, d} of Fig.~\ref{Be8_C12_O16_Ne20}
we show the energy versus projection time $t = L_ta_t$ for $^{8}$Be, $^{12}$C,
$^{16}$O, and $^{20}$Ne respectively for the
LO
interaction A (blue plus signs and dashed lines), LO interaction B (red squares
and dashed lines), LO + Coulomb interaction A (blue crosses and solid lines),
and  LO + Coulomb
interaction B (red triangles and solid lines).  The error bars indicate one
standard deviation errors from the stochastic noise of the Monte Carlo simulations,
and
the lines are extrapolations to infinite projection time using
the ansatz in Eq.~(\ref{E_extrap}).
The results for the ground state energies
are shown in Table~1.

For both interactions A and B, the auxiliary-field Monte Carlo simulations
presented here have far milder Monte Carlo sign cancellations than in previous
lattice simulations of the same systems \cite{Lahde:2013uqa}.  This very
promising  development will allow for much larger
and previously difficult simulations in the future.  The savings come from
two innovations.  The first is the introduction of the nonlocal interactions
in $V_{\rm
NL}$.  Ironically, the implementation of general nonlocal interactions in
quantum Monte Carlo simulations have long been problematic due to sign oscillations.
However, the auxiliary-field implementation of the interactions in $V_{\rm
NL}$ are extremely favorable from the point of view of sign oscillations.
 The reason for this is the very simple structure of the terms in $V_{\rm
NL}$.  This leads to fewer issues with so-called interference sign problems
as discussed in Ref.~\cite{Lahde:2015ona}.  The other
innovation reducing the sign problem is the introduction of the parameter
$b_{\pi}$ in the one-pion exchange interaction.  This decreases the short-distance
repulsion in the $S$-wave channels responsible for some sign oscillations.

\subsection*{Adiabatic projection method}
The adiabatic projection method is a general framework that produces a low-energy
effective theory for
clusters of particles which becomes exact in the limit of large projection
time.
 The details of the methods used here were discussed in Ref.~\cite{Elhatisari:2015iga}, and we review some of the main features
here.
On our $L^3$ periodic 
spatial lattice we consider a set of initial two-alpha states $|{\bf R}\rangle$
labeled
by the spatial separation vector {\bf R}. For the alpha-alpha scattering
calculations presented here we use $L=16\;{\rm fm}$. The initial alpha wave
functions are Gaussian
wave packets which, for large $|{\bf R}|,$ factorize as a product of two
individual alpha clusters,
\begin{equation}
|{\bf R}\rangle=\sum_{{\bf r}} |{\bf r}+{\bf R}\rangle_1\otimes|{\bf r}\rangle_2.
\label{eqn:single_clusters}\\ 
\end{equation}
   The summation over $\bf {r}$ is required to produce states with total
momentum equal to zero. As we have done in Eq.~(\ref{spherical}) for nucleon-nucleon
scattering, we project onto spherical harmonics
$Y_{\ell,\ell_z}$ with angular momentum quantum numbers $\ell,\ell_z,$
\begin{equation}
|R\rangle^{\ell,\ell_z} = \sum_{{\bf R'}}Y_{\ell,\ell_z}(\hat{\bf R'})\delta_{R,|{\bf
R'}|}|{\bf R'}\rangle.
\end{equation} We only consider values for $|{\bf R}|$ less than $L/2$.

The next step is to multiply by powers of the transfer matrix\ to form dressed
cluster
states that approximately span the set of low-energy alpha-alpha scattering
states in our periodic box. We start with the approximate transfer matrix
$M_*$ as in Eq.~(\ref{mstar2}), and then follow with powers of the leading-order
transfer matrix $M$.  After $n_t$ time steps, we have the dressed cluster
states 
\begin{equation}
\vert R\rangle^{\ell,\ell_z}_{n_t} = M^{n_t}M_*^{L'_t}|R\rangle^{\ell,\ell_z}.
\end{equation}The dressed cluster states are then used to compute matrix
elements of the transfer matrix $M$,
\begin{equation}
\left[M_{n_t}\right]^{\ell,\ell_z}_{R',R} =\ ^{\ell,\ell_z}_{\!\!\!\!\!\quad{n_t}}\langle
R'\vert M
\vert R\rangle^{\ell,\ell_z}_{n_t}.\label{Hmatrix}
\end{equation}
 Since the states are not orthogonal, we compute a norm
matrix 
\begin{equation}
\left[N_{n_t}\right]^{\ell,\ell_z}_{R',R} =\ ^{\ell,\ell_z}_{\!\!\!\!\!\quad{n_t}}\langle
R'\vert R\rangle^{\ell,\ell_z}_{n_t}. \label{eqn:norm}
\end{equation}
The radial adiabatic transfer matrix is defined as the matrix product,
\begin{equation}
\left[ {M^a_{n_t}} \right]^{\ell,\ell_z}_{R',R} = 
\left[N_{n_t}^{-\frac{1}{2}}M_{n_t}
N_{n_t}^{-\frac{1}{2}} \right]^{\ell,\ell_z}_{R',R}.
\label{eqn:Adiabatic-Hamiltonian}
\end{equation}
Just as we have done for nucleon-nucleon scattering, we impose a spherical
hard wall boundary at some radius $R_W$.  For large $n_t$ the standing waves
of the radial adiabatic transfer matrix are used to determine the elastic
phase shifts for alpha-alpha scattering.
As explained in Ref.~\cite{Elhatisari:2015iga}, this scattering calculation is
extended out  to very large volumes of $L^3 = (120\;{\rm fm})^3$ using single
alpha-particle simulations and including long-range Coulomb interactions
between the otherwise non-interacting alpha particles at large distances.

\begin{figure}[!ht]
\centering
\caption{Alpha-alpha \textit{S}-wave extrapolations for interaction
A. LO + Coulomb results (circles)
for the $S$-wave phase
shift for interaction A at several laboratory energies versus the number
of time steps $L_t=2n_t+1$. The panels {\bf a}, {\bf b}, {\bf c},
{\bf d}, {\bf e}, {\bf f}, {\bf g} correspond to laboratory energies $E_{\rm
Lab}$~=~1.0, 2.0, 3.0, 4.5, 6.5, 8.5, 10.0~MeV respectively. The error bars
indicate
one standard deviation uncertainty due
to Monte Carlo errors. The dot-dashed lines show the extrapolation
to the $L_t \rightarrow \infty$ limit, and the hatched regions show
the one standard deviation error estimate for the extrapolation. \bigskip}
\includegraphics[scale=1.2]{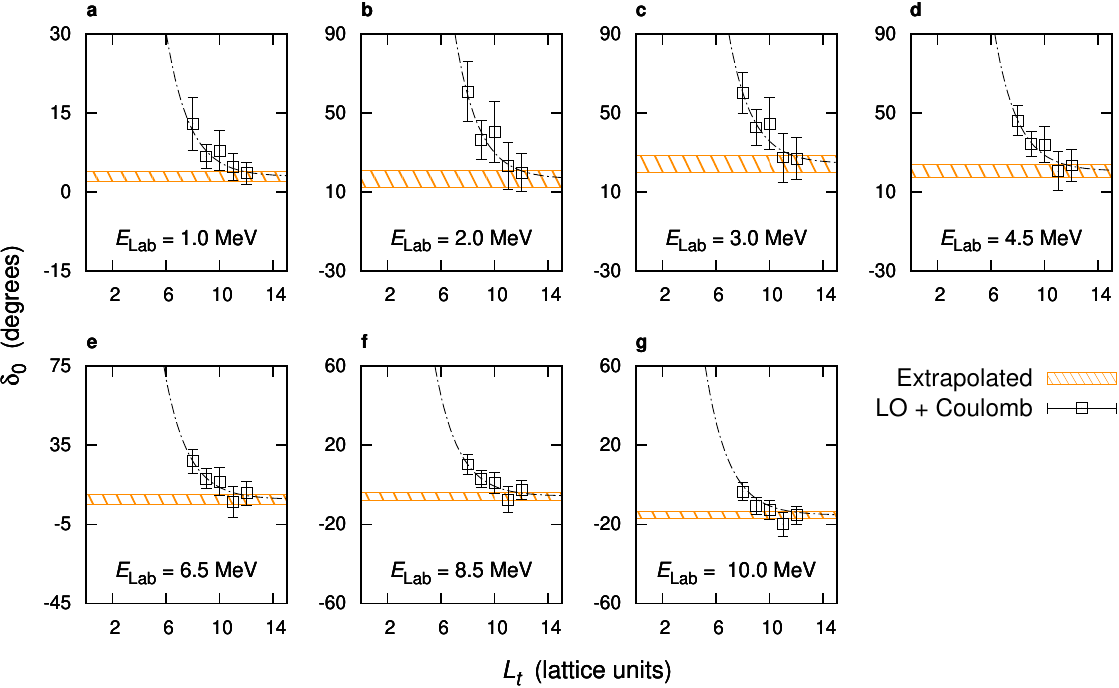}
\label{A_extrapolation}
\end{figure}

\begin{figure}[!ht]
\centering
\caption{Alpha-alpha \textit{S}-wave extrapolations for interaction
B. LO + Coulomb results (circles)
for the $S$-wave phase
shift for interaction B at several laboratory energies versus the number
of time steps $L_t=2n_t+1$. The panels {\bf a},
{\bf b}, {\bf c},
{\bf d}, {\bf e}, {\bf f}, {\bf g} correspond to laboratory energies $E_{\rm
Lab}$~=~1.0, 2.0, 3.0, 4.5, 6.5, 8.5, 10.0~MeV respectively. The error bars
indicate
one standard deviation uncertainty due
to Monte Carlo errors. The dot-dashed lines show the extrapolation
to the $L_t \rightarrow \infty$ limit, and the hatched regions show
the one standard deviation  error estimate for the extrapolation. \bigskip}
\includegraphics[scale=1.2]{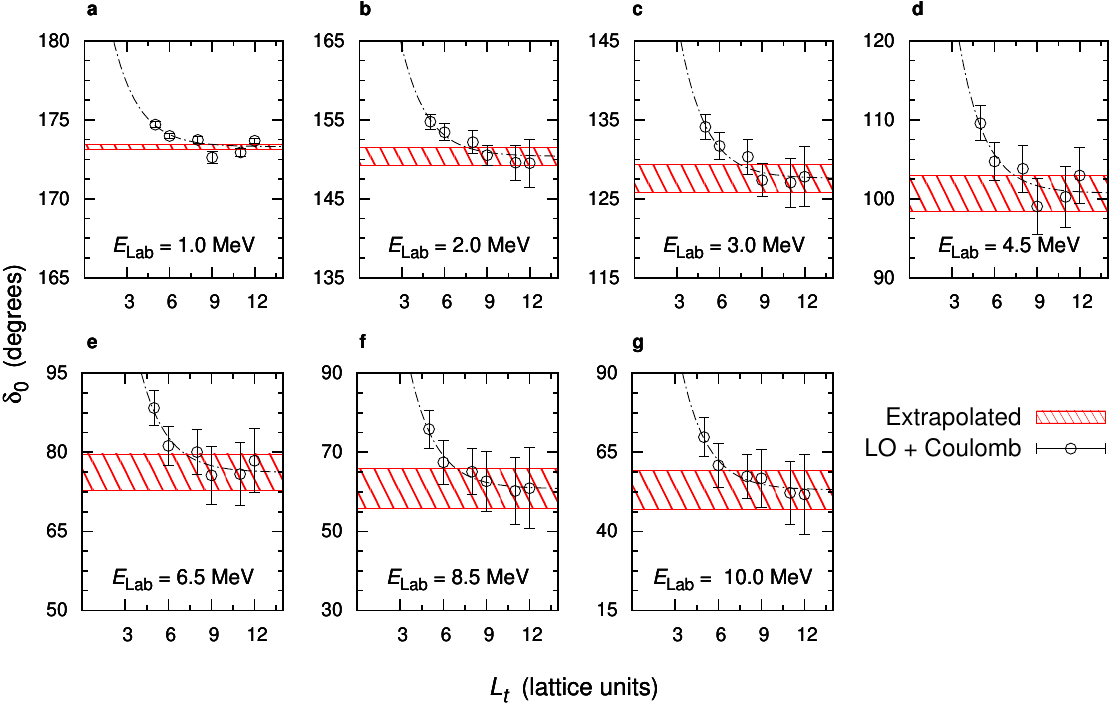}
\label{B_extrapolation}
\end{figure}

In Fig.~\ref{A_extrapolation} we plot the LO + Coulomb $S$-wave
phase
shifts for interaction A at several laboratory energies versus the number
of time steps $L_t=2n_t+1$. The analogous LO + Coulomb $S$-wave phase shifts
for interaction B are shown in Fig.~\ref{B_extrapolation}.
For both of these figures, the panels {\bf a}, {\bf b}, {\bf c},
{\bf d}, {\bf e}, {\bf f}, {\bf g} correspond to laboratory energies $E_{\rm
Lab}$~=~1.0, 2.0, 3.0, 4.5, 6.5, 8.5, 10.0~MeV respectively. The
error bars indicate
one standard deviation uncertainties due
to Monte Carlo errors, and the dot-dashed lines show the extrapolation
curve for the $L_t \rightarrow \infty$ limit.
We use the ansatz\begin{equation}
\delta_0(L_t,E) = \delta_0(E) + c_0(E)\exp[-\Delta E \, L_t a_t],
\end{equation}
 where $\delta_0(E)$ is the extrapolated phase shift.  The hatched regions
in Fig.~\ref{A_extrapolation} and \ref{B_extrapolation} show
the one standard deviation error estimate of the extrapolation.

\subsection*{Tight-binding approximation and potential}

The tight-binding approximation is a simple qualitative picture where the
alpha particle is treated as a compact object with a small but nonzero radius,
$R_{\alpha}$.  As the name suggests, it is conceptually similar to the tight-binding
model of electronic structure commonly used in condensed matter physics.
Here we provide some further details of the direct and exchange terms in the calculation of the tight-binding potential between two alpha clusters. Let us consider a nucleon-nucleon interaction
in continuous space of the form\begin{equation}
\frac{1}{2}\int d^3{\bf R}d^3{\bf r'}d^3{\bf r}\; V_{s_4,i_4;s_3,i_3}^{s_2,i_2;s_1,i_1}({\bf
r'},{\bf r})a^\dagger_{s_3,i_3}({\bf R}-{\bf r'}/2)a^\dagger_{s_4,i_4}({\bf
R}+{\bf r'}/2)
a_{s_2,i_2}({\bf R}+{\bf r}/2)a_{s_1,i_1}({\bf R}-{\bf r}/2).
\end{equation}
The indices $s_1,s_2,s_3,s_4$ correspond to spin, while $i_1,i_2,i_3,i_4$
correspond to isospin.  For $r~>~R_{\alpha}$, the tight-binding potential
$V_{\rm TB}(r)$ can be divided into two contributions, 
\begin{equation}
V_{\rm TB}(r) =V^{\rm direct}_{\rm TB}(r)+ V^{\rm exchange}_{\rm TB}(r),
\end{equation}
where the direct term is
\begin{equation}
V^{\rm direct}_{\rm TB}(|{\bf r}|) = \sum_{s_{24},{i_{24}}}\sum_{s_{13},{i_{13}}}
\int_{|{\bf r'}-{\bf r}|<R_{\alpha}} d^3{\bf r'} \; V_{s_{24},i_{24};s_{13},i_{13}}^{s_{24},i_{24};s_{13},i_{13}}({\bf
r'},{\bf r}),
\end{equation}
and the exchange term is
\begin{equation}
V^{\rm exchange}_{\rm TB}(|{\bf r}|) = -\sum_{s_{23},{i_{23}}}\sum_{s_{14},{i_{14}}}
\int_{|{\bf r'}+{\bf r}|<R_{\alpha}} d^3{\bf r'} \; V_{s_{14},i_{14};s_{23},i_{23}}^{s_{23},i_{23};s_{14},i_{14}}({\bf
r'},{\bf r}).
\end{equation}

\subsection*{Ground state energies as a function of $\lambda$}
\begin{figure}[!ht]
\centering
\caption {Ground state energies versus $\lambda$. We plot the LO ground state energies $E_{^4{\rm He}}$, 1/2\;$E_{^8{\rm Be}}$, 1/3\;$E_{^{12}{\rm C}}$,
1/4\;$E_{^{16}{\rm O}}$, 1/5\;$E_{^{20}{\rm Ne}}$ versus the parameter $\lambda$ which interpolates between $V_{\rm A}$ and $V_{\rm B}$.}
\includegraphics[scale=0.6,angle=0]{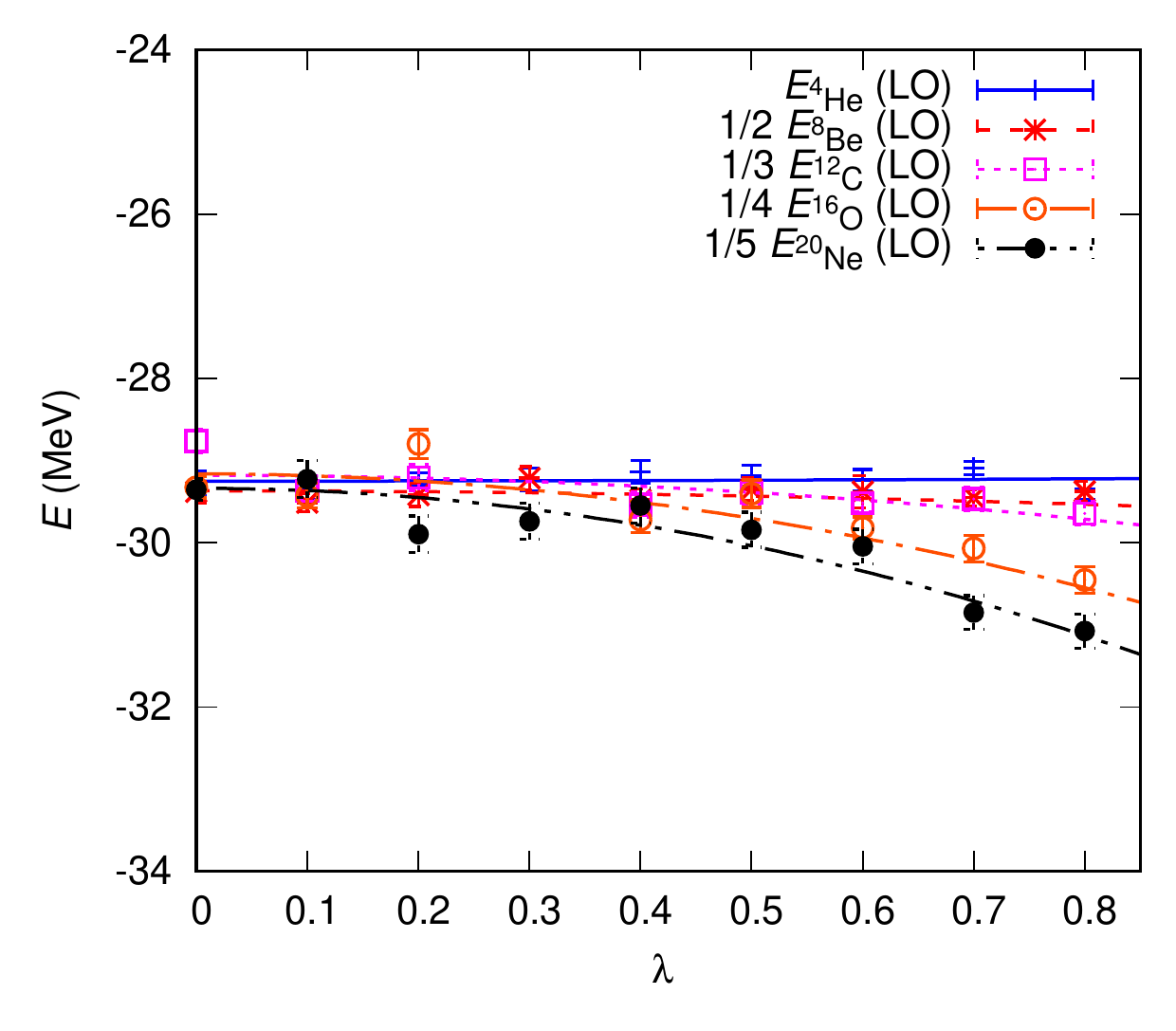}
\label{lambda_rescaled}
\end{figure}

We consider the one-parameter family of interactions, $V_{\lambda}=(1-\lambda)V_{\rm
A} + \lambda V_{\rm B}$ with the Coulomb interactions switched off.  At the phase transition point the alpha clusters become non-interacting in the dilute limit, and so we should find the following simple relationship among the ground state energies provided that the finite volume is sufficiently large:
\begin{equation}
E_{^4{\rm He}} = 1/2\;E_{^8{\rm Be}} = 1/3\;E_{^{12}{\rm C}} = 1/4\;E_{^{16}{\rm
O}} = 1/5\;E_{^{20}{\rm Ne}}.
\end{equation}
In Fig.~\ref{lambda_rescaled} we plot the LO ground
state energies $E_{^4{\rm He}}$, 1/2\;$E_{^8{\rm Be}}$, 1/3\;$E_{^{12}{\rm C}}$,
1/4\;$E_{^{16}{\rm O}}$, 1/5\;$E_{^{20}{\rm Ne}}$ versus $\lambda$.  We see that the phase transition occurs at $\lambda_{\infty} = 0.0(1)$.

To determine the critical point $\lambda_{20}$ when $^{20}$Ne becomes bound, we compare
$E_{^{20}{\rm
Ne}}$ with the threshold energy $E_{^{16}{\rm O}}$ + $E_{^4{\rm He}}$.  For this analysis we also include the finite-volume energy one obtains at infinite $S$-wave scattering length for the $^{16}$O + $^4$He system. At infinite scattering length the energy of any two-body system with reduced mass $\mu$ in a periodic box
of size $L$ is \cite{Lee:2005fk,Beane:2003da}
\begin{equation}
\Delta E=\frac{4\pi^2d_1}{mL^2},
\end{equation}
where
\begin{equation}
d_1 \approx -0.095901.
\end{equation}
We find that the critical point for the binding of $^{20}$Ne is $\lambda_{20}$~=~$0.2(1)$.  A similar analysis for the binding of the other alpha nuclei finds  $\lambda_{16}$~=~$0.2(1)$ for $^{16}$O, $\lambda_{12}$~=~$0.3(1)$ for
$^{12}$C, and $\lambda_{8}$~=~$0.7(1)$ for
$^{8}$Be.
\subsection*{Code Availability}
All codes used in this work are freely available and can be obtained by contacting
the authors.

\bibliography{References}

\bibliographystyle{apsrev}

\newpage

\end{document}